# Polyphonic Pitch Tracking with Deep Layered Learning

Anders Elowsson
anderselowsson@gmail.com

*Abstract*—This paper presents a polyphonic pitch tracking system able to extract both framewise and note-based estimates from audio. The system uses several artificial neural networks in a deep layered learning setup. First, cascading networks are applied to a spectrogram for framewise fundamental frequency ($f_0$) estimation. A sparse receptive field is learned by the first network and then used as a filter kernel for parameter sharing throughout the system. The $f_0$ activations are connected across time to extract pitch contours. These contours define a framework within which subsequent networks perform onset and offset detection, operating across both time and smaller pitch fluctuations at the same time. As input, the networks use, e.g., variations of latent representations from the $f_0$ estimation network. Finally, incorrect tentative notes are removed one by one in an iterative procedure that allows a network to classify notes within an accurate context. The system was evaluated on four public test sets: MAPS, Bach10, TRIOS, and the MIREX Woodwind quintet, and performed state-of-the-art results for all four datasets. It performs well across all subtasks: $f_0$, pitched onset, and pitched offset tracking.

## I. INTRODUCTION

THIS article presents a novel method for estimating multiple frame-level fundamental frequencies ($f_0$s) and transcribing the onsets and offsets of notes in polyphonic music. These tasks are among the most studied in the research field of music information retrieval (MIR).

Each pitched tone generally consists of periodic waves (partials) with frequencies at integer multiples. In Western polyphonic music many such tones sound simultaneously, most of them related through simple integer ratios (e.g., 4/3). This means that partials of different tones will overlap extensively. Variations in, e.g.: timbre of the performing instruments, audio processing (e.g., equalization), and microphone characteristics [1], further complicates the multiple $f_0$ estimation task.

The most basic parameters of music notes are the onset time and offset time, together with the associated pitch, transcribed in the *note tracking* task. During note tracking, the system must try to account for the same uncertainties as for $f_0$ estimation. In addition, the envelope and timbre vary considerably at note starts for different instruments, further complicating the task.

### A. Fundamental frequency estimation

Attempts at $f_0$ estimation date back to the 1970s [2]. Prominent directions have been to: utilize the mismatch between detected spectral peaks and ideal harmonics [3-6]; assume spectral smoothness of partials [4, 7-10]; decompose spectral frames using a set of basis patterns with non-negative matrix factorization (NMF) [11-14], prior subspace analysis (PSA) [15], or probabilistic latent component analysis (PLCA) [16-18]; or to learn compositional layers with unsupervised learning [19].

Early supervised learning implementations used separate support vector machine (SVM) models for each note pitch [20, 21]. By applying a convolutional neural network (CNN) that operate across the pitch direction in spectrograms with logarithmically spaced frequency bins, pitch-invariant processing can instead be achieved. A CNN with small local receptive fields was proposed in [22], and kernels with a range of approximately an octave were used in [23]. A recent system used time, frequency, and a third dimension consisting of the first five harmonics and a subharmonic [24]. On a similar theme, magnitudes of the first partials have been used as basic features for classification [25]. Another recent system used a CNN in combination with a bidirectional long short-term memory (LSTM) network to compute framewise predictions [26].

### B. Note tracking

Different methodologies for detecting discrete note events have been proposed. Many of these implementations use frame-level predictions of pitch salience as input, trying to discretize this information across time into a set of notes. This has been done with, e.g., dynamic Bayesian networks [27], or a hidden Markov model (HMM) [20, 21, 28] coupled with a musicological model [29, 30]. Other musicological models have been proposed by Sigtia et al. [22, 31]. When relying on frame-level predictions, the reliability can be increased by median filtering across time [14, 32], and by removing thresholded $f_0$ activations shorter than 30-80 ms [16-18, 33].

In many implementations, note onsets are extracted separately. Early systems using neural networks for pitched onset detection were proposed by Marolt [34, 35]. In these systems, a separate network was trained for each note pitch (i.e., no weight sharing across pitch), using input from framewise partial tracking networks. Some systems extract input data guided by detected $f_0$ activations, for example, the estimated note length from frame-level pitch-blocks [36], the magnitude of pitch activation-changes [37], or features capturing the spectral flux [38] for a log-frequency spectrogram filtered by a morphological dilation. The latter technique is used to account for pitch-shifts and has been described for onset detection as the *harmonic onset* function [39] or *vibrato suppression* [40]. Attack-decay NMF models have been used for piano transcription [41, 42]. Other methods have estimated note starts directly in piano recordings, using an LTSM network [43], or cascading classifiers [44]; but an LSTM network has also been used in conjunction with framewise estimates [26].

Offset detection is a subject that has received far less attention than onset detection. Offsets have often been extracted together with onsets, e.g., from thresholded $f_0$ activations [16-18, 33]. Dressler [45] tracked variations in partial magnitudes detecting offsets when the level drops, weighted according to the stability of the level across time. Similar principles have been used for monophonic transcription [46].

*C. Overview of the article*

This article presents a "deep layered learning" (DLL) [47] methodology for polyphonic transcription. Intermediate targets are used to perform a "structural disentanglement," i.e., the inherent organization of the music is extracted at intermediate layers of processing. One purpose is to facilitate invariant processing of the data. In the case of pitched onset and offset detection, this is achieved by extracting pitch contours, and applying neural networks operating across the contours to compute activations. Section II gives an extended overview of the system and describes the neural networks. The initial spectrogram filtering is described in Section III. Then, detailed descriptions of the learning steps are made in Sections IV-VIII and the method used for regularization presented in Section IX. The training and test sets are described in Section X, together with a description of the evaluation methodology. Section XI presents the evaluation of the system on four public datasets and Section XII offers conclusions and discussion, including a further discussion of DLL in the context of this publication.

## II. METHOD OVERVIEW

This section gives an overview of the system, and Sections III-VIII provides a more detailed description of each processing step. The video-graphical abstract included as supplementary material provides additional visualization of the processing.

*A. System overview*

Fig. 1 shows a flowchart of the system. It contains six learning steps (blue arrows in the figure), where the first ($N_1$) performs logistic regression and the following ($N_2$-$N_6$) are neural networks. Each network uses representations (green boxes) computed in earlier networks. Between each network, the output activations are processed (yellow boxes) to prepare the system for the next learning stage in the processing chain. Finally, the output estimates are produced (red boxes). The networks of the system were trained in an iterative fashion, (1) processing the training tracks up until a new network to compute its input, (2) training the network, (3) processing the training tracks with the newly trained network, (4) taking the output from the network and processing the tracks further to prepare the input of the next network, etc. References to Fig. 1 are indicated in *italics* in the following overview.

First, the variable-Q transform (*VQT*) [48] was applied to the audio files, which produced a spectrogram with spectral bins at a log-frequency distribution. The spectrogram was filtered by applying a noise floor that whitens the spectrum and ensures signal level invariance, resulting in the spectrogram *L* (Section III). The first learning step ($N_1$) then computed tentative framewise $f_0$ estimates from a linear combination of 50 spectral bins from *L* (Section IV). The best performing combinations of spectral bins were selected by iteratively adding bins in a forward selection. The 50 bins can be understood as a sparse learned kernel operating across frequency. Prominent tentative $f_0$ activations were extracted with *peak-picking* from the *Tentogram* – the output from the linear combination of spectral bins across all pitches. The extracted estimates were updated by performing a more accurate estimation with a bigger network ($N_2$), resulting in a *Pitchogram* (Section V). Computed $f_0$ activations form regions in the Pitchogram that expand over time. These regions, corresponding to pitch *contour*s, were extracted.

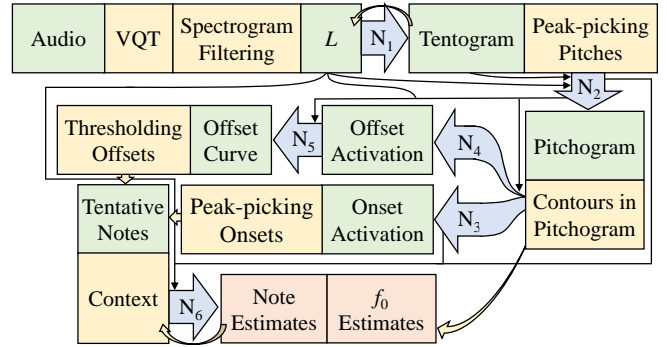

Fig 1. Overview of the polyphonic pitch tracking system. Representations (green), intermediate processing steps (yellow), the six learning steps (blue arrows – $N_1$-$N_6$), and the final note and $f_0$ estimates (red). Representations (including latent such) passed from one network to the input of another ("skip connections") are indicated by black arrows.

The variation over time in the activations of the last hidden layer of $N_2$, as well as other relevant spectral features and output activations, were extracted along the pitch contours. They were then used as input to a network ($N_3$) computing an *onset activation* curve for each contour (Section VI). After smoothing the onset activation, tentative onsets were extracted by *peak-picking*. The same input was then used to compute an *offset activation* curve with a fourth network ($N_4$), from which the final offset position was determined by the fifth network $N_5$ (Section VII). *Tentative notes* of each ME were then established from the estimated onsets and offsets.

Finally, each of these notes was evaluated in the note network ($N_6$), using previously computed representations as well as information about neighboring notes (pitch and time) to compute the probability that a note is correct (Section VIII). Incorrect tentative notes were removed in an iterative procedure, one by one for the ME, starting with the tentative notes that are most likely to be incorrect. The remaining notes correspond to the *note estimates*, and the pitch ridges from onsets to offsets were used as $f_0$ *estimates* (both evaluated in Section XI).

*B. Network overview*

Table 1 shows an overview of the learning steps in the system. The networks used the filtered spectrogram (*L*) together with information from earlier networks to achieve their respective objective. Networks that fed their hidden layers as representations to later networks had two hidden layers. All instances were feedforward neural networks, and all but the first so-called multilayer perceptrons (MLPs). As expanded on below, they however exhibit similar parameter sharing as in CNNs due to the local scope of each training example. The networks used the scaled conjugate gradient training method and were trained with batch gradient descent using the cross-entropy cost function. All networks were trained with early validation stopping. The number of allowed failed epochs before stopping was lower in earlier networks, to discourage too pronounced overfitting to the training set at an early stage. The input features of $N_2$-$N_6$ were normalized so that the minimum and maximum value across the training set fit the range between -1 and +1. The non-linearities for the hidden layer(s) were hyperbolic tangent (tanh) units and for the output, the sigmoid function was used.

| $N_x$ | Objective | Input | Network output | Size |
|---|---|---|---|---|
| $N_1$ | Detect tentative $f_0$s ($t_0$s) | $L$ | Tentogram | 65 |
| $N_2$ | Classify $t_0$s (i.e., define $f_0$s) | $L$, $N_1$ | Pitchogram | 176-100-14 |
| $N_3$ | Detect onsets in pitch contour | $L$, $N_2$ | Onset activation | 1485-50-30 |
| $N_4$ | Detect offsets in pitch contour | $L$, $N_2$ | Offset activation | 1485-50-30 |
| $N_5$ | Determine offset position | $L$, $N_2$, $N_4$ | Offset curve | 253-100 |
| $N_6$ | Classify notes | $L$, $N_2$, $N_3$ | Note classification | 2893-150 |

Table 1. Networks used by the system, including their abbreviation, objective, input and output representations, and size (excluding the output layer of size 1).

An overview of the run-time operation is provided in Fig. 2. The figure shows how the filtered spectrogram (green) is transformed by $N_1$ into the Tentogram (orange) containing tentative $f_0$s. Detected tentative $f_0$s are then filtered by $N_2$ using two hidden layers (white) to produce a Pitchogram (orange). A contour is illustrated in dark orange, and the networks $N_3$-$N_5$ operate across this contour to detect onsets and offsets, which define the start and end of a tentative note illustrated in red. Features from the note are used to make the final classification in $N_6$.

Although the networks were implemented as MLPs, at run-time their first layer acts like a sparse convolutional kernel operating across pitch/frequency ($N_1$-$N_2$), or time ($N_3$-$N_5$). This is due to the single target annotation (local scope) used for each training example, where the input is extracted as a sparse vector with the same shape for all training examples – but shifted according to the corresponding pitch and time of the annotation. The number of hidden neurons in the following hidden layer therefore corresponds to the number of kernels and the depth of the output in Fig. 2. Continuing the analogy, subsequent layers of each network use 1×1 convolutional kernels extending across depth (see small square kernels in Fig. 2), with the number of kernels corresponding to the depth of the subsequent layer. Although the note classification network ($N_6$) also use inputs that are extracted across pitch and time, it receives input from so many representations that it instead was illustrated as an MLP in Fig. 2.

III. COMPUTING A FILTERED SPECTROGRAM

A. Computing the initial spectrogram

The initial spectrogram was computed for each ME with the VQT [48], using 60 bins per octave, with a minimum frequency computed from MIDI pitch 26 (around 37 Hz) and a maximum frequency of 14.45 kHz. The parameter that controls the trade-off between time and frequency resolution in the lower frequencies was $\gamma = 11.6$. By using $\gamma$ above 0, the time resolution is increased in the lower frequencies at the cost of a lower frequency resolution, which is beneficial for, e.g., onset detection. The resulting magnitude spectrogram $M$ had 518 log-frequency spaced bins. The specific settings were used since they give a hop size very close to 128 samples/frame. The hop size was then adjusted to be 256 samples/frame (about 5.8 milliseconds, $ms$) by removing roughly every other frame.

B. Filtering the spectrogram

The spectrogram was filtered to compensate for the time-varying sound level (at a phrase resolution) and the long-term cumulative average spectrum of the recording. These aspects of the audio vary depending on, for example, instrumentation and the dynamics of the performance [49]. By compensating for the time-varying sound level, the same processing weights may be applied when estimating $f_0$s at soft and loud sections, thereby providing invariance to the level. By compensating for the long-term spectrum, the same processing weights may be applied when, for example, processing tracks played by instruments that vary in spectral characteristics such as the violin or piano, thereby providing some invariance to instrumentation. These *spectral whitening* procedures are rather common for pitch tracking [e.g., 50]. An overview of the filtering process and the resulting spectrogram is provided in Fig. 3. The two-dimensional matrix $V^s$ was computed to represent the long-term average spectrum (pane 2), and the 1-dimensional vector $V^l$ was computed to represent the time-varying sound level (pane 3). These were combined to form the matrix $V^{ls}$ (pane 4), the time-varying average spectrum. By subtracting $V^{ls}$ from the initial spectrogram (Section III-A, pane 1), the spectrogram $L$ could finally be computed (pane 5). Details for the computation of $V^l$, $V^s$, and $V^{ls}$ are provided in the Appendix.

The filtered spectrogram had three different dynamic ranges:

$$L = \max\{(20 \times \log_{10} M) - V^{ls}, 0\}, \quad (1)$$

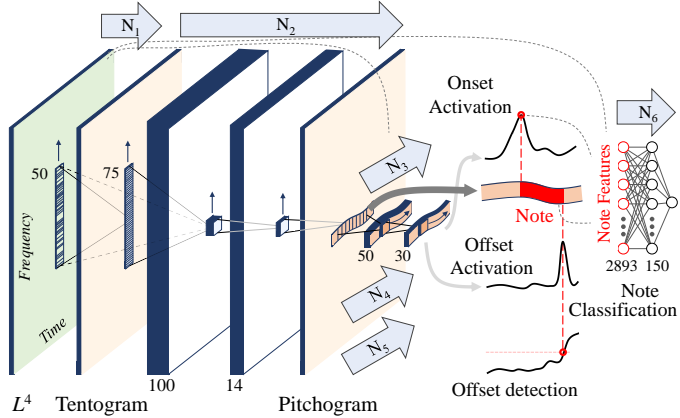

Fig. 2. An overview of the networks used by the system at run-time, illustrating the processing steps as a combination of CNNs and MLPs. A sparse kernel convolves the spectrogram representation $L^4$, producing a Tentogram consisting of tentative $f_0$s. A deeper and bigger network ($N_2$) is applied to tentative $f_0$s, using a similar kernel strategy for its first layer. An identified contour in the resulting Pitchogram is used as a framework by networks $N_3$-$N_5$ when computing onsets and offset avtivations. A detected onset and offset for a note is illustrated with red dots in the activation curves (black lines). It is shown how these points bound the extension of a note (red) across time. Spectral features are finally gathered across the extension of the note together with context information, and these collected note features are used for the note classification.

$$L_{15+} = \max\{(20 \times \log_{10} M) - V^{ls} + 15, 0\}, \quad (2)$$
$$L_{25+} = \max\{(20 \times \log_{10} M) - V^{ls} + 25, 0\}. \quad (3)$$

In the system, $L$ was used for $f_0$ estimation, $L_{15+}$ for onset detection, and $L_{25+}$ for note classification. The reason for initially using $L$, which may discard relevant information, is that, for computational purposes, a linear combination of spectrogram levels is used to compute the Tentogram. Therefore, the values should reflect the relative importance of the respective magnitudes in the spectrogram. The reason for using $L_{15+}$ and then $L_{25+}$ is that these spectrograms will contain more information at faint signal levels (which will be rather noisy). After filtering, the frequency dimension of $L$ was upsampled to four times its original size with linear interpolation; referred to as $L^4$.

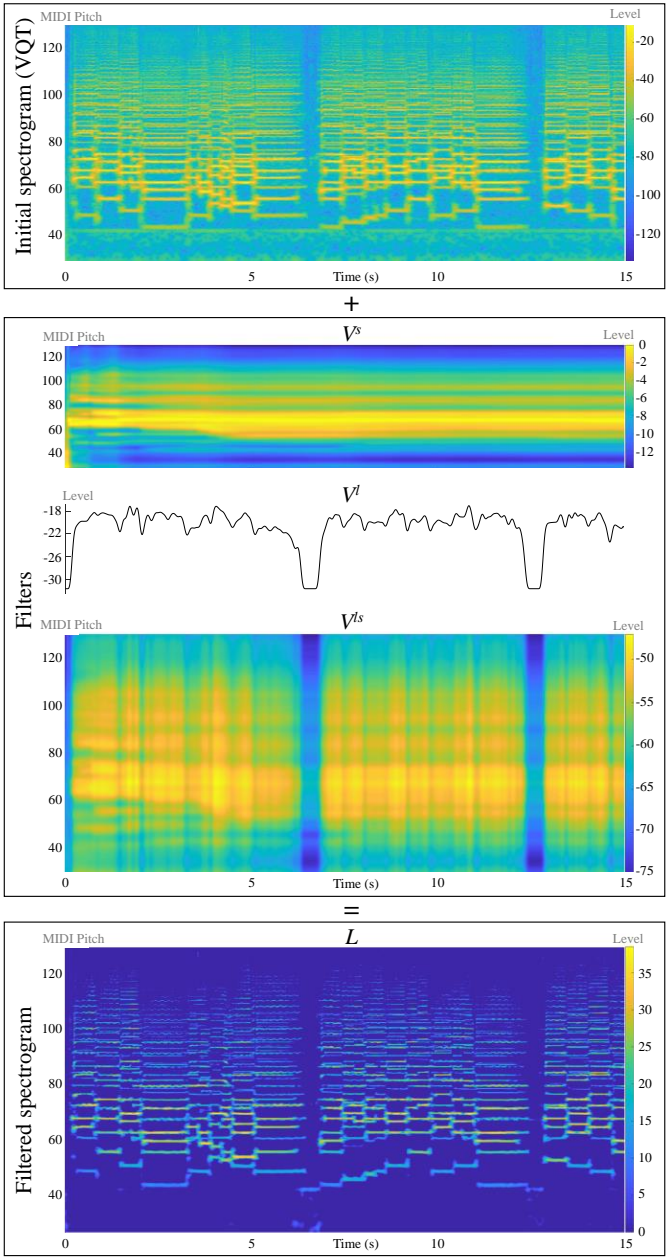

Fig. 3. The initial VQT spectrogram, the filters $V^s$ and $V^l$ combined into the filter $V^{ls}$, and the output of the filtering, $L$; shown for 15 seconds of ME No 1 of the Bach10 test set.

## IV. TENTOGRAM

### A. Overview

To account for time-varying tonal information in a pitch tracking model, $f_0$s need to be computed with a sufficient resolution both across time and frequency. There is a significant computational cost associated with this if the processing is to have the necessary complexity. Therefore, the pitch detection step was divided into two sub-steps in the system. First, tentative pitches were computed using a fast spectral summation to produce a Tentogram as described in this section, and then detected tentative $f_0$-candidates were further evaluated (to produce a Pitchogram, as described in Section V).

The Tentogram network ($N_1$) was trained to estimate framewise $f_0$s from spectral energies in $L^4$, using a pitch kernel with bins sparsely distributed across frequency (Section IV-C). The distribution of these bins was initially learned by the network, and the network can be described as a single sparse convolutional kernel. However, additionally, a set of 15 DCT components (Section IV-E), defined for each $f_0$ training example (Section IV-B), were used to perform a Tentogram whitening. Weights were inferred by a neural network using no hidden layers[1], and these were used to specify a computationally efficient run-time operation (the spectral summation described in Section IV-F) that produces the final Tentogram. The parameter sharing mechanism of the kernel extends across, nearly, the full spectrum. In image processing, parameters are often shared for processing adjacent pixels. Auditory perception of music however differs in that, for example, octave-spaced frequency bins are more closely related.

### B. Selecting $f_0$ training examples

Training was repeated two times, with a slightly different way of selecting training examples for the first and second iteration. In the first iteration, the annotated $f_0$s ($T_A$) were used for the true class in the network classification, and the false class $T_F$ consisted of training examples at $\pm\{3\ 4\ 5\ 6\ 7\ 8\ 9\ 12\ 19\ 24\}$ semitones relative to each annotated true $f_0$. These positions were an initial rough selection of pitches that can produce erroneous $f_0$ estimates (e.g., octave errors). Any $T_F$ in a time frame within 50 cents of a $T_A$ was removed. After training the kernel in the first iteration, the system was run with the trained kernel, and an output Tentogram computed for each ME. The computed tentative pitches were then used to select additional and refined training examples. In this case, for each annotated $f_0$, the closest pitch detected in the Tentogram was used to replace the actual annotated pitch, if a relevant pitch (within 50 cents) could be found. This is useful, as vibratos or other smaller pitch shifts are not included in the annotations generated from the MIDI files. The same technique was used to improve the pitch of the false training examples at the previously specified pitches. Furthermore, any additional incorrect pitches detected in the Tentogram were added to $T_F$ as annotations in the false class, so that these could be suppressed as well.

### C. Learning active bin indices of the pitch kernel

Tentative pitch activations were computed from a weighted sum of signal levels at sparsely spaced frequencies in $L^4$. Which frequencies could be the most useful in this regard? Relevant frequencies are arguably those that correspond to pitch partials of an evaluated $f_0$, as well as other frequencies that can be used for indicating that a certain pitch is incorrect. These *other frequencies* will be referred to as *inter-partials*, positioned in between the partials and also positioned at subharmonics of an evaluated $f_0$. Relevant inter-partials can be hard to determine analytically, and they also vary with musical styles, harmonic structure, instrumentation, and the noise profile of the audio recordings. It is very likely that a manual selection would lead to mistakes and the inter-partial positions were therefore instead learned by the system. Given the run-time behavior of $N_1$, the system can be conceived of as using a *learned sparse partial receptive field* or *pitch kernel*. This kernel was then used by many learning modules in the system.

---
[1]This is very similar to logistic regression, but a neural network implementation with a sigmoid activation function was used for computational reasons.

Various properties of the kernel $K$ will be denoted by subscripts. The bin indices (relative to the evaluated pitch bin) selected to be active in the sparse kernel will be referred to as $K_j$. These consists of both partial and inter-partials positions. The extracted spectrogram levels (from $L^4$) at these indices will be referred to as $K_L$. The vectors $K_j$ and $K_L$ will always have the same $K_{size}$, which refers to the number of elements in the vectors. Since the same pitch kernel was used across all $f_0$s, $K_j$ is a fixed set of indices (relative to the processed $f_0$), whereas $K_L$ will vary across pitch and time. The first 11 partials of the $f_0$s were initially added to the kernel ($K_{size} = 11$). To do this, the bin indices of the $n$th partial was computed from

$$\lfloor \log_2(n) \times bpo \times uf \rfloor, \quad (4)$$

where $bpo$ represents the number of bins per octave of the VQT (60), and $uf$ represents the upsampling factor (4). The operation $\lfloor x \rfloor$ is used for rounding the result to the closest integer. An important benefit of using the upsampled spectrogram $L^4$ is the reduction of these rounding errors, which ensures that partials and learned inter-partials are added with a high precision.

Inter-partials were then chosen by forward selection, iteratively adding each inter-partial to $K_j$, evaluating the performance with $K_L$ as input to the network, and finally choosing the inter-partial that improved performance the most after all had been evaluated. Inter-partials were selected from –40 to +45 semitones relative to the $f_0$. The resolution was 20 bins/semitone, so the number of evaluated inter-partials in each iteration of the forward selection procedure was $85 \times 20 + 1 = 1701$ (discounting $K_{size}$).

To reduce the computational cost, each iteration was performed in two separate steps – a fast step using fewer training examples, and a second step for the most relevant inter-partials with more training examples. In the first step, every 2900$^{th}$ time frame in the training set was used for training, and every 500$^{th}$ frame of the validation set was used for validation. The network was trained with early stopping and the performance for each variation of $K$ calculated. The result was added to a performance vector $X$, of length 1701. Values at bin indices already added to $K_j$, and therefore not evaluated, were filled in by interpolation of neighboring values. The vector $X$ was then smoothed with a Hann window of size 13 (to reduce variability), and the 45 lowest values (best performance) further evaluated in the second step. For the second step, every 100$^{th}$ frame of the training and validation set was used, and the best performing of the 45 evaluated bin indices added permanently to $K_j$. The forward selection was halted automatically at $K_{size} = 50$.

After the pitch kernel had been defined, the network was finally trained with more data, using every other time frame from the training and validation set. The bias term determined for the network during training will be denoted $K_b$, and the computed weights $K_W$.

### D. Bin distribution of the learned kernel

The 50 bin indices of the pitch kernel (11 pre-determined, 39 determined during training) are shown in Fig. 4 and provided together with the learned weights in the Appendix. They range from bin -705 to 874, with 17 of the learned indices being negative and 22 positive. The distribution indicates that, excluding pre-determined partial positions, bins below the $f_0$ were roughly as important as bins above the $f_0$ for performing pitch tracking.

There seems to be an interesting pattern for the learned part of the kernel; the network repeatedly selected bin indices at a frequency ratio relative to the $f_0$ corresponding to fractions of small prime numbers. There were 13 prime fractions within 15 cents of the ideal bin index for a prime pair. These bin indices may be well suited to uniquely signify other simultaneous tones, due to the structure of Western tone scales and the integer ratios of partial components. Neighboring bins were selected by the system at four frequencies; and each time, one of the bins was assigned a relatively large positive weight and one a relatively large negative weight. These bin-pairs, therefore, act as an edge detector on the spectrogram. The effect is that they can suppress or increase the likelihood of an $f_0$ based on the slope of the spectrogram level (across frequency).

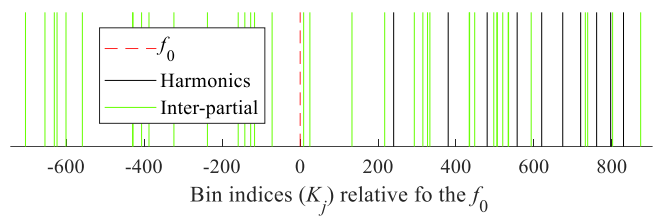

Fig. 4. The bin indices ($K_j$) used by the pitch kernel, at a resolution of 5 cents/bin. The partial positions ($f_0$ = red dashed line, harmonics = black lines) were pre-specified, and the inter-partial positions (green lines) were learned during training.

### E. Tentogram whitening with DCT components

The value of 15 DCT components, representing the pitch of the processed $f_0$, were also supplied as input. This enabled the network to whiten the Tentogram using a compressed representation. The solution solves two problems:

1. It can be assumed that the linear combination will produce more false activations at some pitch ranges than others. For example, harmonics at the higher pitch range may not always be sufficiently suppressed.
2. The pitch of the $f_0$ cannot express anything but a low/high pitch preference, due to the linear $N_1$. The first components

of the DCT can, however, express preference at a smoothness varying with the number of included components.

A precomputed basis function $B_{DCT}$ of size $15 \times 1563$ was formed by the first 15 components of the DCT III (excluding the constant DC component). For each training example, the index position $i_p$ into $B_{DCT}$ was derived from the MIDI pitch $P$ according to

$$i_p = (P - 25.8) \times 20 \qquad 1 \leq i_p \leq 1563, \qquad (5)$$

and a vector, $X_{DCT}$, of 15 components extracted. The vector $X_{DCT}$ was included as input to the network, both during the kernel forward selection and the final training. The whitening vector was finally derived by matrix multiplication

$$W = W_{DCT} \times B_{DCT}, \qquad (6)$$

where $W_{DCT}$ corresponds to the computed weights for $X_{DCT}$ from the network. The whitening vector, shown in Fig. 5, was added to each frame of the Tentogram as specified in Section IV-F.

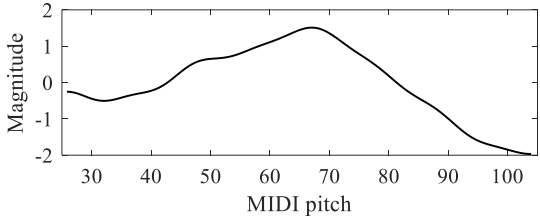

Fig. 5. The Tentogram whitening vector $W$ computed during training. The system learned to give lower weight to high pitches, and higher weight to two octaves around MIDI pitch 65.

### F. Run-time

At run-time, the weighted sum of $K_L$ was computed with a fast spectral summation, by summing $K_{Size}$ copies of the 2-dimensional spectrogram $L^4$. First, each copy was shifted $-1 \times K_j$ frequency bins, bins outside of MIDI pitch 25.85-103.95 (a range of 1563 bins in $L^4$) were discarded, and potentially missing rows of bins in a copy padded with zeros. This resulted in the 3-dimensional matrix $L_K^4$ with the size $K_{Size} \times 1563 \times nf$, where the number of time frames in the ME is represented by $nf$. Each copy was then weighted according to its computed weight $K_W$, and the copies summed. This operation was done by matrix multiplication

$$\text{Tentogram} = K_W \times L_K^4, \qquad (7)$$

where $K_W$ had the size $1 \times K_{Size}$. The processing generates a tentative $f_0$ activation for every pitch bin in a spectrogram-like representation, and this representation is therefore called a Tentogram. Using matrix multiplication to compute the Tentogram ensures computational efficiency. The Tentogram was however further refined. The bias from the network training ($K_b$) and the whitening vector $W$ were summed into a vector $c$

$$c = W + K_b + 3.5, \qquad (8)$$

which was added to each time frame, and all values below 0 in the Tentogram were then set to 0. The addition of 3.5 ensures that rather improbable tentative $f_0$s also are processed by the subsequent pitch network. The Tentogram was finally filtered with a 2-dimensional Gaussian $\sigma = 3$, truncated to the size of $11 \times 11$ bins (64 ms $\times$ 0.55 semitones).

## V. PITCHOGRAM

### A. Extracting tentative $f_0$s ($t_0$s) for classification

Pitch peaks were detected in the Tentogram ($T$), for each time frame $i$, as the pitch bins $j$ where $T_{i,j} > T_{i,j-1}$ and $T_{i,j} > T_{i,j+1}$. By using parabolic interpolation, the pitch of each $t_0$ was estimated to a resolution of 1 cent. Detected peaks defined tentative $f_0$s ($t_0$s), and these $t_0$s were classified in a second, larger network ($N_2$), in order to produce a refined Tentogram – the Pitchogram. Each $t_0$ was first evaluated as *true* or *false* by the evaluation procedure presented in Section X-C. Any $t_0$s that had been evaluated as false but that were within 50 cents of an annotated $f_0$ were discarded from training.

### B. Network

Fig. 6 gives an overview of the Pitchogram network ($N_2$) used for classifying the $t_0$s. The network had two hidden layers (size 100 and 14). This meant that various network sizes could be tested while keeping the last hidden layer fixed in size (which is useful since subsequent networks will use hidden layer activations from $N_2$).

As input for the classification of each $t_0$, $K_L$, previously used for training the Tentogram network (Section IV-C), was extracted by taking a cross-section of $L_K^4$ at the corresponding frame $i$ and bin $j$ of the $t_0$. Additionally, a "pooled" (maximum magnitude across frequency) version of the same features was used. The idea behind this procedure was to account for variation, while also implicitly giving a hint to the network of the approximate absolute distance between existing partial peaks in the spectrogram and partial and inter-partial frequencies defined in the pitch kernel[2]. The maximum level was extracted within $\pm$ 6 bins in $L_K^4$, corresponding to $\pm$ 30 cents.

Furthermore, the maximum $t_0$-peaks were extracted at 73 locations in the Tentogram, $\pm$ 36 semitones relative to each evaluated $t_0$. One value was computed for each semitone taking the maximum $t_0$ at $\pm$ 50 cents. For semitones where no $t_0$ was detected, the value 0 was used. The sums of all $t_0$s $>$ 36 semitones and all $t_0$s $<$ -36 semitones were used as two additional input features. Finally, the pitch of the processed $t_0$ was also added as an input feature.

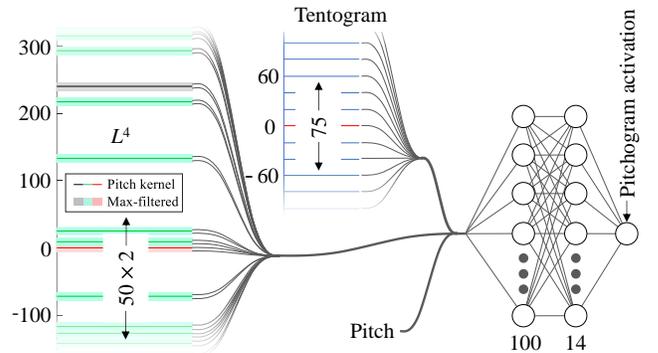

Fig. 6. The input to $N_2$, used to compute the pitchogram activation for a $t_0$. Red lines indicate the $t_0$. For the $L^4$ spectrogram, black is the pre-defined position of the first harmonic, and the learned inter-partials in $K_j$ are indicated by green. For the Tentogram, blue is the semitone offsets at which the maximum activation of close $t_0$s were extracted.

---

[2]If there is a partial peak close to the kernel frequency, the difference between the pooled and non-pooled signal level gives an indication of the absolute frequency distance between the kernel frequency and that partial peak.

## C. Run-time

Due to the input structure, the network at run-time can roughly be understood as an initial layer with 100 convolutional filters, followed by 14 1x1 convolutional filters having a depth of 100 (see $N_2$ in Fig. 2 of the overview), and a final layer of 1 1x1 convolutional filter with depth 14. An activation was computed for each $t_0$ from the trained network. The value prior to the sigmoid activation function of the output layer was used, and the constant 3.6 added. This value was selected under the assumption that activations for rather unlikely $f_0$s could also provide valuable information. All activations below zero were discarded, and the remaining activations inserted in the Pitchogram (frequency resolution: 100 cents/semitone) at the pitch-bin corresponding to the previously interpolated pitch (Section V-A). The Pitchogram was then smoothed across pitch with a Hann window of size 41.

## D. Output

To illustrate and evaluate the performance at this stage, the threshold was computed that maximized the framewise F-measure ($\mathcal{F}_{fr}$) (see Section X-C) for the *validation set* (evaluating all possible thresholds in steps of 0.01). The computed threshold, 3.87, was higher than the zero-threshold used during run-time.

Fig. 7 shows the spectrogram $L$, the Tentogram computed from $L$, and the Pitchogram (with the optimized threshold), for 6 seconds of audio. As shown, the 2-step process gradually refines the estimates while upsampling the resolution, unveiling, for instance, vibratos. There are very few activations outside annotated notes, and all annotated notes have a corresponding activation.

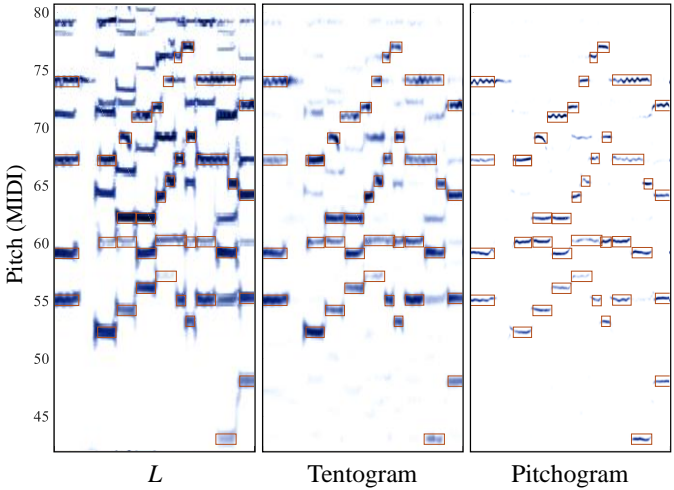

Fig. 7. The filtered spectrogram representations $L$, computed Tentogram, and the Pitchogram with an optimized threshold, for 6 seconds of audio (time points 18 to 24) in Bach10 ME No 1. Ground truth annotations based on onset and offset times are marked with a red rectangle (the annotations are oftentimes slightly incorrect). Zoom in to review how vibrato features are unveiled.

## E. Detecting contours in the Pitchogram

Regions of $f_0$ activations were extracted from the Pitchogram to facilitate further processing. This was done by identifying connected activations above 0. A connectivity of eight was used – each above-zero pixel in the image could be connected to all eight surrounding pixels. In image processing, similar types of regions have often been referred to as "blobs." Related, a system for audio-to-score alignment [51] relied on blob-detection from pitch-salience estimates, using start-points and end-points of the blobs as onsets and offsets.

Subsequently, regions with a similar pitch that follow in close succession were merged to be able to track noisy $f_0$ activations over a longer period of time. The pitch of each region was first computed as the weighted centroid of its $f_0$ activations. Then, for each region, ordered based on start time, any new region with a pitch within 50 cents of the old region and a start time between 0 and 130 ms after the old region was merged to form a combined larger region. The merging was done in an iterative procedure, updating the pitch, onsets, and offsets of regions after each merging.

In the next step, a peak ridge was extracted along time for each region, corresponding to the pitch contour. The bin indices of the ridge were defined from the maximum $f_0$ activation in each frame of the region. If no pitch activation was present for a specific time frame (due to merged regions), the bin index was determined by linear interpolation between detected indices before and after the evaluated time frame. The contour ridge was also extended backward 30 frames before the first frame of the region, to fully capture potential onsets at the start of the region. These frames were assigned the same pitch bin index as the first frame (nearest neighbor extrapolation). Fig. 8 shows the pitch contour ridge (red line) extracted from a merged region and a non-merged region. The pitch contour ridge extends horizontally at the edges of the two regions (extrapolation) and connects regions at edges during interpolation. The abrupt incline upwards at 5.7 seconds is a result of a spurious small (erroneous) region being merged in between bigger regions.

Finally, the pitch activation was computed with $N_2$ for all frames of the contour ridge where it had not previously been computed (e.g., at extrapolated time frames). After this, each region had been assigned a one-dimensional contour across time, which was used to compute an onset and offset activation curve. As the contour traces the pitch fluctuations, the subsequent processing will benefit from an accurate and high pitch resolution and activation, while being invariant to local pitch variations of the tone. This *"tone-shift-invariant"* (or *"vibrato-invariant"*) property of the system promotes generalization across instrumentation – differences in tone stability are negated when tone features at each time frame are extracted relative to the established contour ridge.

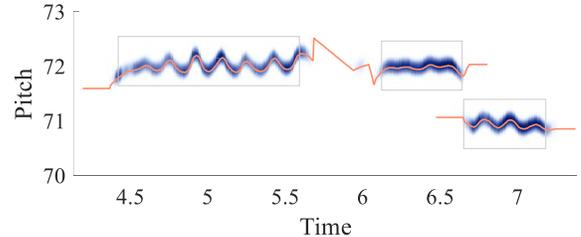

Fig. 8. Merged pitch regions (MIDI pitch 72, left) and a single region (MIDI pitch 71, right), with the corresponding pitch contour (red line) from which onsets and offsets were computed. The notes were played by the violin and are from the four-voiced ME No 2 of the Bach10 test set. Ground truth annotations are marked with a grey box.

## VI. Onset Detection

An onset activation function was computed with a neural network (the onset network $N_3$) along each pitch contour. The input to the network was mainly the change in activation or signal level across time for previously computed representations. This included the *neural flux*, the *spectral flux*, and the absolute *pitch flux*, all described in the following subsections, with sampling indices (corresponding to filter sizes) provided in the Appendix. Computing changes in activations or the spectrum provides an additional layer of invariance. If the values at which these changes happen are less relevant than the magnitude of the changes, this pre-processing can increase performance, as it relieves the network from inferring this connection across the whole input vector. However, the real values were also provided at a single time frame, to provide the network with more complete information. The input structure is described for each time frame of a contour, and it was extracted in the same way for all frames of the contour.

### A. Neural flux

The preceding pitch network was trained to activate at $f_0$s between the start and end of each tone. Pitch activations along contours will therefore generally rise at the start of new notes. These pitch activations were included as input to the onset network can infer the presence of a note onset from them.

Ideally, the activations from the output layer of the $f_0$-network would be enough to identify onsets. However, accurately detecting all present fundamental frequencies can be very hard, not only for computer models but also for skilled listeners; especially in dense polyphonic textures. It can therefore be expected that output activations from $N_2$ are insufficient to detect onsets accurately. As listeners get the benefit of tracking music along time, they may perceive pitch onsets by relying on various clues, for example regarding how harmonics vary in loudness across time. Clues of these types can also be expected to be encoded deeper in the layer-wise structure of $N_2$. Therefore, fluctuations in the activations of the 14 neurons in the last hidden layer of $N_2$ were also used as input to $N_3$. This will be referred to as the "neural flux." Fig. 9 illustrates how the activations from different time frames of a single neuron in the last hidden layer of $N_2$ are sent to the neural flux function. This function computes a vector containing differences in the activations, which is sent to the input of $N_3$.

### B. Spectral flux

Although the neural flux should capture many of the relevant changes in the spectral characteristics that can be associated with note onsets, the objective of $N_2$ was never to track these onsets. In many instruments, the spectral characteristics vary considerably during an onset, and the spectrum is not the same as in the subsequent steady part of a note. Therefore, a more low-level representation was also supplied to the network, in the form of the spectral flux, computed from $L_{+15}$.

### C. Additional features

The amplitude of contour pitch variations was included as input, computed as the absolute values of the derivative (changes) in the contour pitch across time. The intuition for computing the absolute values is that information about the *presence of a fluctuating pitch* is much more valuable for the task than information about the direction of potential pitch fluctuations. The derivative of the loudness variation ($V^l$) computed in the preprocessing of the spectrogram was also included.

Three additional single features were provided; the local time index in the region, the distance to the end of the region, and the mean pitch of the region.

### D. Training

The onset network was trained to classify whether a certain time frame of a contour corresponded to a note start or not. The contour frames annotated as true (1) for training fulfilled the following criteria:

- identical time as the onset time of an annotation (when quantized to 256 samples),
- a pitch within 55 cents of that annotation.

The surrounding time frames ($\pm$ 7 frames) in the contour were discarded from training, and the rest of the frames were set to false (0). To obtain a more even balance between true and false annotations during training, about 95 % of the training examples annotated as false were excluded from training; these were chosen randomly.

### E. Peak-picking onsets

The neural network was applied to every frame of the contours, producing an onset activation curve (*OC*) for each contour. The sigmoid output activation function used during training of the network was excluded during run-time. The *OC* was then filtered, using four parameters determined in a grid search on the training and validation set, as described in the Appendix. The onset curve after filtering will be referred to as *OCS*, and it is shown in red in Fig. 10 in Section VII. Peaks in the *OCS* defined tentative onsets. Exact onset times were determined with parabolic interpolation, using neighboring values in the *OCS*.

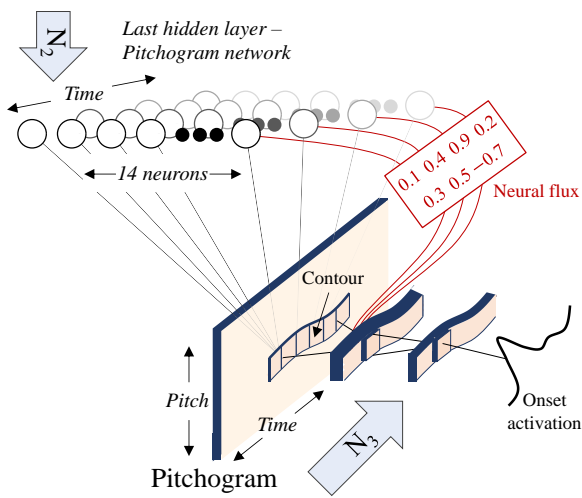

Fig. 9. Illustration of how activations in $N_2$ are used for onset detection in $N_3$. The last hidden layer of $N_2$ consists of 14 neurons (some omitted) and is illustrated for 4 different time frames. The hidden layers of different time frames have been connected since their corresponding output activations belong to the same contour. The fluctuation in activation (neural flux) from each of the 14 neurons is computed across time and used as input for computing an onset activation in $N_3$, as illustrated by red for one specific neuron across 4 time frames.

## VII. Offset Detection

For each tentatively detected onset, an offset position was also computed. This was done in two stages. First, the same features and network size used for onset detection were also used for a network ($N_4$) producing an offset curve (*OFC*). Offset positions for each region were annotated by using a Hann window with a width of 13 frames and a maximum amplitude of 1, centered at the annotated offset position. This design was chosen due to the higher uncertainty in timing for offset positions.

A straightforward solution would have been to then choose offsets based on a peak-picking on the *OFC*. However, this approach could lead to missed offset for tones with slowly decreasing sound levels (i.e., hammered or plucked instruments) or tones in dense polyphonic textures. Therefore, an offset detection network ($N_5$) was incorporated in a second stage. For the annotations of this stage, the frames before the annotated offset for a detected note were set to 0, and frames after the annotated offset were set to 1. During five frames centered at the offset position, they linearly increased from 0 to 1. Features accumulated from the start of the tone were then supplied to the network during training. These included activations from $N_3$, the *OFC*, and the pitch kernel signal levels ($K_L$). Further details are provided in the Appendix. The desired outcome was for the network to take accumulated evidence of an offset into account, with its output activation reaching 0.5 close to the annotated offset, subsequently increasing.

Shown in Fig. 10 is the Pitchogram representation, and the $f_0$, onset and offset activations for a note played by the violin with vibrato in an excerpt of the four-voiced Bach10 dataset.

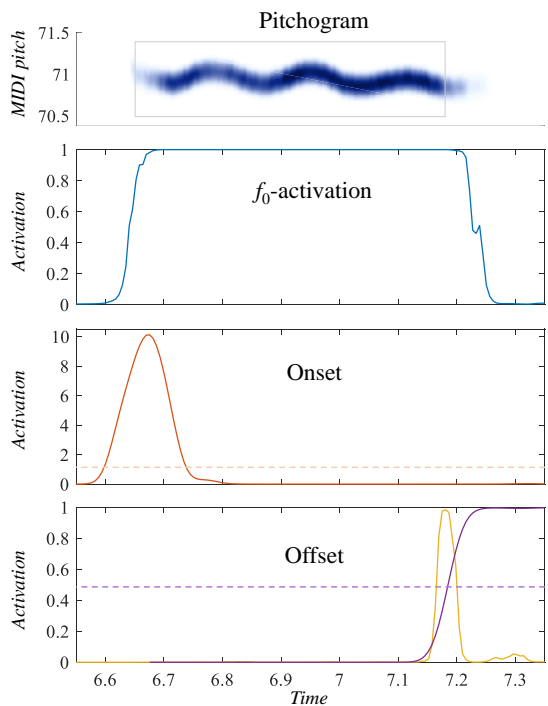

Fig. 10. The Pitchogram representation of a note (Me No 2 of Bach10) with the note annotation (grey box), its $f_0$ activation after the sigmoid activation function (blue line), and output activations for onsets and offsets. The onset activation (red line) is the *OCS* prior to thresholding at 1.2. The offset activation, *OFC* (yellow line), is shown together with the smoothed offset detection activation (purple line), and its corresponding threshold (dashed purple line).

## VIII. Note Classification

Finally, notes were processed by the note classification neural network ($N_6$), and each note assigned a probability of being correct. The input features focused on context, providing information about the associated region, spectral composition across the note, and onset characteristics.

### A. Input features

Given that the onset network was trained mostly based on the change of activation/signal level, and that this information, therefore, should be encoded in the resulting *OC*, the input to the note network instead mostly consisted of raw activations/signal levels. These were provided from the *OC*, the last hidden layer of $N_3$, the signal levels (relative to the note contour). Furthermore, the tentative note was provided with a set of context features where it was compared with other concurrent tentative notes. Further details are provided in the Appendix.

### B. Run-time operation

During run-time (after training the network), notes were removed iteratively, in ascending order according to their computed probability of being correct. During this process, after each note-removal, the input features and note probabilities of surrounding notes affected by the presence of the removed note were updated. This included a new calculation of the offset for any note directly preceding the removed note in the same region. The procedure is used to provide an accurate context during classification. For example, when a correct offset has been determined for a note, after another onset that previously shortened the note has been removed, it can be expected that note features computed across its duration become more accurate.

### C. Training

Training was done in two steps. In both steps, a set of training examples (*Tr*) was created for each ME in the training and validation set, and the note network $N_6$ trained to classify each example as a correct or incorrect note. First, the system was run up until the note-removal step for each ME and all extracted tentative notes evaluated, creating *Tr*. $N_6$ was then trained to classify the notes in *Tr*.

In the second step, the system was run through the note-removal step by using the version of $N_6$ trained during the first step. Both the kept and removed notes were gathered as tentative training examples $Tr_T$. The removed notes had their input features extracted at the time of their removal in the first step. The note evaluation was performed on $Tr_T$, and the note that had been removed first during run-time was added to the empty set *Tr*, and removed from $Tr_T$. This procedure was repeated for all notes removed during run-time. If a removed note was evaluated as true, the evaluation of notes in close proximity (pitch and time) was re-done, as the removal could have altered their evaluation from false to true. Then, the notes kept during run-time were also added to *Tr*, using their assigned evaluation, and a new version of $N_6$ finally trained with the notes in *Tr* as training examples.

## D. Computing a framewise $f_0$ estimate

A framewise $f_0$ estimate was generated from the transcribed notes. The $f_0$-predictions were defined as the time-varying pitch for each note (along with the contour), for all time frames between the onset and the offset. Then, the combined predictions for each time frame were analyzed. For all framewise predictions within 60 cents, the prediction with the highest $f_0$ activation was kept and the other predictions removed. This method gave slightly better framewise predictions (signified by a higher $\mathcal{F}$-measure) than the thresholding described in Section V-D.

## IX. REGULARIZATION

When using the same training set for each step of the deep layered learning process, the system will become over-reliant on the activations extracted from previous layers. It will appear as if the extracted activations are very predictive for the next step, when, in fact, information from previous representations may be more useful. The potential for overfitting to earlier network outputs will increase at every learning step of the system. For regularization, the frequency response of the initial spectrograms $M$ was therefore varied randomly for all tracks in the training set between training each learning module, through filtering. Details concerning the creation of these filters are provided in the Appendix. Four of the randomly created frequency filters are shown (in the equivalent decibel scale) in Fig 11. The regularization technique was added after first having run the complete training and testing without it, thereby providing an informal test of its effect. After training with regularization, $\mathcal{F}$ (Section X-C) decreased by around 0.5 for the training set and increased by around 1 for the test sets.

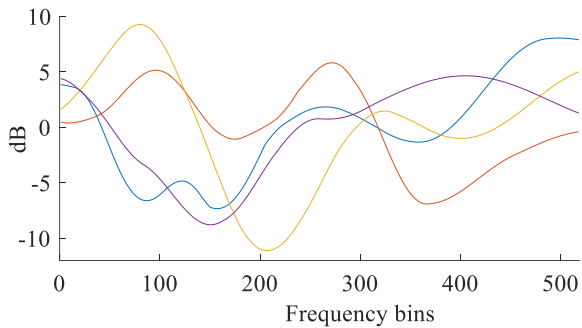

Fig. 11. Four randomly generated frequency filters used for regularization.

## X. DATASETS AND EVALUATION METHOD

### A. Training sets

Two training sets were generated from MIDI files, using the GeneralUser GS SoundFont[3]. The audio samples were analyzed to increase the accuracy of the corresponding onset and offset annotations in the *training set*, as described in the Appendix.

Each MIDI file was subsequently preprocessed. First, pedaling was removed, instead extending the offsets for notes sustained by the pedal. For sustained instruments, onsets immediately following an offset with *the same pitch* were detected, and a gap inserted by shortening the first MIDI note. The length of the gap $x$ was adjusted based on the estimated offset delay $d$ of the instrument generating the audio according to $= 0.7d$, restricting $x$ to the range 0.02-0.3 s. This ensured that the second onset was audible. Notes with an offset within 30 *ms* of the onset time (according to the accurate, analyzed, onset and offset times) were extended to 30 *ms*. If this shifted an annotated offset time to happen after the next onset of *the same pitch*, the note was removed from the MIDI file.

The first dataset, "*Attacked*," consisted of plucked and hammered instruments and was based on 145 MIDI files of classical music collected from the Disklavier e-competition[4]. The dataset was initially screened to remove any of the compositions also present in the MAPS and Bach10 datasets (Section X-B) used for testing. No music by Mozart was included in the training set, thereby increasing train/test independence concerning composition style. MIDI files longer than three minutes were split into three-minute excerpts to facilitate a more efficient load balancing during training, resulting in 454 MIDI files. The instrument used to synthesize each MIDI file was selected randomly according to the following probabilities for each GM instrument: Pianos including harpsichord and clavinet (GMI 1-8), 9.5 %. Chromatic percussion (GMI 9, 12-13), 2.3 %. Acoustic guitars (GMI 25-26) 3.5 %. Electric guitars (GMI 27-28, 31), 3.3 %.

The second dataset (*Sustained*) consisted of sustained instruments. The dataset was generated from 369 MIDI files of Bach chorales which had previously been derived from MusicXML files created by Margaret Greentree[5]. Overlapping compositions present in the Bach10 dataset were removed. Audio samples from five instrument groups were used to generate the audio files; Organ (GMI 17-19, 21-24), Violin (GMI 41-45, 47, 49-52), Brass (GMI 57-62, 64) Reed (GMI 65-72), and Pipe (GMI 73, 75-77, 79-80). For each MIDI file, five versions were made, resulting in a total of 1845 musical excerpts (MEs). Each version had four voices. The GMI for each voice was selected randomly, with the only restriction that each ME should contain four different instrument groups. Furthermore, for each of the five versions, the global tempo was varied by a factor 0.9-1.15, and the pitch transposed $\pm$ 2 semitones. The onset time of each note was also varied slightly.

The total number of onsets/offsets and the total number of framewise $f_0$s of each dataset are presented in Table 2. About 10 % of the generated MEs were selected for the validation set. During training, the datasets were merged into one training and one validation set.

| Number of | *Attacked* | *Sustained* |
|---|---|---|
| **Framewise $f_0$s** | $1.03 \times 10^8$ | $1.00 \times 10^8$ |
| **Onsets/offsets** | $5.89 \times 10^5$ | $4.78 \times 10^5$ |

Table 2. The two datasets consisting of attacked or sustained instruments. The datasets were designed to have roughly the same number of annotations for the tasks.

### B. Test sets

Four test sets were used. The first was the *Bach10* dataset[6] [5], which consists of four-voiced Bach chorales. Each voice had been recorded separately (real instrument performances) with a bassoon, a tenor saxophone, a clarinet, and a violin and merged into a monaural polyphonic ME. The second test set was the *MAPS* dataset[7] [10], consisting of piano recordings generated from MIDI using seven virtual pianos (software) and an upright Yamaha Disklavier piano recorded with both close and ambient microphones. This dataset consists of 270 MEs, created from 160 classical compositions. The third set was the *TRIOS* dataset[8], which consists of five chamber music trio MEs. Each voice was recorded separately (real instrument performances) and the annotations manually aligned. The fourth dataset was just a single track, a *Woodwind* recording of Beethoven's Variations for String Quartet Op.18 No. 5 [52]. Each of the five voices was recorded separately and then combined in a monaural polyphonic audio file.

---

[3] http://schristiancollins.com/generaluser.php
[4] http://www.piano-e-competition.com/
[5] Previously hosted at www.jsbchorales.net
[6] http://www2.ece.rochester.edu/~zduan/resource/Bach10%20Dataset_v1.0.pdf

### C. Evaluation methodology

The performance of the system was evaluated considering framewise $f_0$ estimation, note tracking (onsets) and offset detection. When computing evaluation metrics, each estimate and annotation was first assigned a true or false label. A maximum deviation of 50 cents was allowed for both framewise and note-based estimates (the estimated note pitch was used for both onsets and offsets). For note transcription, the onsets were allowed a maximum deviation of 50 ms; for offset transcription, 100 ms was used. A metric considering both onsets and offsets was also computed, using 50 ms for onsets as before, and either 50 ms or 20 % of the note length for offsets – whichever was the largest. The methodology for pairing annotations and estimates proposed in [53] was used. This matching strategy will produce slightly fewer matched pairs (and therefore slightly lower results) than algorithms evaluated by the Python mir_eval library [54]. Matched pairs were used to compute the precision $\mathcal{P}$ (the proportion of estimates that were correct), and recall $\mathcal{R}$ (the proportion of annotations with a matching estimate). Finally, the F-measure, $\mathcal{F}$, was computed:

$$\mathcal{F} = 2 \frac{\mathcal{P} \times \mathcal{R}}{\mathcal{P} + \mathcal{R}}. \quad (9)$$

The F-measure will be denoted $\mathcal{F}_{fr}$ for framewise estimates, $\mathcal{F}_{on}$ for onsets, $\mathcal{F}_{off}$ for offsets, and $\mathcal{F}_{off}^{on}$ for both onsets and offsets. Some previous studies instead use accuracy $A$, so that was also computed as described by [53]. As other datasets use "percentages" instead of fractions, the computed metrics $\times$ 100 will be used throughout the article. When not otherwise stated, the datasets were evaluated based on the average performance of all individual notes or framewise $f_0$s within the dataset.

## XI. EVALUATION

To compare the performance with previous systems evaluated across the four datasets, the systems and abbreviations from [53] were used.

### A. Main results

Table 3 shows the framewise performance, and the performance when requiring both onsets and offsets to be correct, for the four test sets. Table 4 shows the note-based performance, evaluating onsets and offsets separately.

There seems to be a difference in performance for datasets where the system had seen similar types of examples (Bach10

[7] http://www.tsi.telecom-paristech.fr/aao/en/2010/07/08/

and MAPS) and for datasets where no similar examples were present (TRIOS and Woodwind). The system, however, achieves high results for these unfamiliar examples as well. Offsets are harder to estimate than onsets for the system, which can be expected as these are less defined for hammered instruments. However, for datasets only containing sustained instruments (Bach10 and Woodwind), the performance is almost equal (differing measurement windows unaccounted for). Since the onset annotations for the Disklavier subsets of the MAPS dataset and the whole Bach10 dataset have a rather inaccurate timing, the $\mathcal{F}_{on}$ with an 80 ms window is provided here for future reference: MAPS = 91.4, Bach10 = 91.9.

| Test Set | $A_{fr}$ | $\mathcal{F}_{fr}$ | $\mathcal{P}$ | $\mathcal{R}$ | $\mathcal{F}_{off}^{on}$ | $\mathcal{P}$ | $\mathcal{R}$ |
|---|---|---|---|---|---|---|---|
| **Bach10** | 85.6 | **92.2** | 91.4 | 93.1 | 84.2 | 83.0 | 85.4 |
| **MAPS** | 64.5 | **78.4** | 84.9 | 72.8 | 58.0 | 58.7 | 57.3 |
| **TRIOS** | 56.0 | **71.8** | 91.1 | 59.2 | 39.1 | 40.0 | 38.1 |
| **Woodwind** | 57.4 | **72.9** | 74.7 | 71.2 | 51.1 | 54.9 | 47.8 |

Table 3. The framewise performance and the note-based performance requiring both onsets and offsets to be correct, for the four test sets.

| Test Set | $A_{on}$ | $\mathcal{F}_{on}$ | $\mathcal{P}$ | $\mathcal{R}$ | $\mathcal{F}_{off}$ | $\mathcal{P}$ | $\mathcal{R}$ |
|---|---|---|---|---|---|---|---|
| **Bach10** | 79.0 | **88.3** | 87.0 | 89.6 | 88.0 | 86.7 | 89.3 |
| **MAPS** | 82.4 | **90.3** | 91.4 | 89.3 | 70.1 | 71.0 | 69.3 |
| **TRIOS** | 72.1 | **83.8** | 85.8 | 81.8 | 47.0 | 48.2 | 45.9 |
| **Woodwind** | 68.9 | **81.6** | 87.7 | 76.3 | 75.8 | 81.5 | 70.8 |

Table 4. The note-based performance of the system concerning onsets and offsets separately for the four test sets.

### B. Bach10

A performance comparison on Bach10 for note tracking and $f_0$ estimation is presented in Fig. 12. The proposed system clearly outperforms other systems for both tasks.

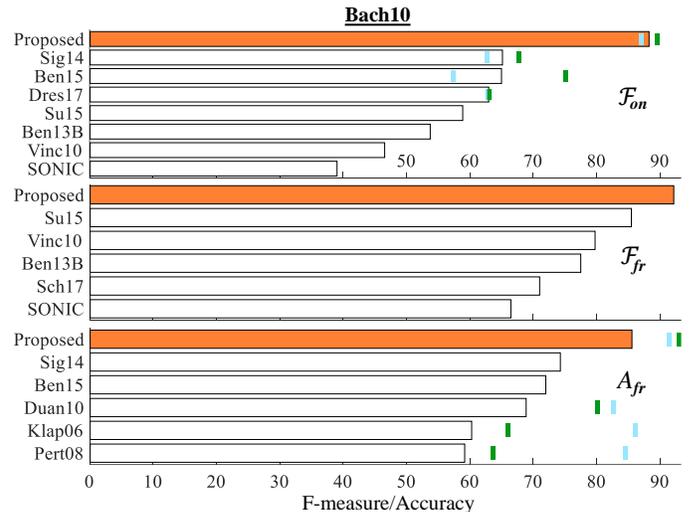

Fig. 12. Comparing $\mathcal{F}_{on}$, $\mathcal{F}_{fr}$ and $A_{fr}$ for the Bach10 test set. Dark green horizontal markers represent $\mathcal{R}$ and light blue $\mathcal{P}$. Abbreviations are from [53].

### C. MAPS

Table 5 shows the framewise performance, and note-based performance for onsets, offsets, and both, for the MAPS dataset broken down into various subsets. The first nine rows of the table are the results for the different pianos, and the last two rows are the combined performance for the Disklavier piano and the 7 synthesized pianos respectively. Many systems only compute performance statistics for the first 30 seconds of the tracks in ENSTDkCl, to save time/memory for resource-intensive systems. These results (ENSTDkCl30) are therefore used for comparison and highlighted in bold in the table.

[8] https://c4dm.eecs.qmul.ac.uk/rdr/handle/123456789/27

| Subset | $\mathcal{F}_{fr}$ | $\mathcal{P}$ | $\mathcal{R}$ | $\mathcal{F}_{on}$ | $\mathcal{P}$ | $\mathcal{R}$ | $\mathcal{F}_{off}$ | $\mathcal{P}$ | $\mathcal{R}$ | $\mathcal{F}_{off}^{on}$ | $\mathcal{P}$ | $\mathcal{R}$ |
|---|---|---|---|---|---|---|---|---|---|---|---|---|
| AkPnBcht | 81.8 | 87.6 | 76.7 | 95.3 | 97.2 | 93.5 | 72.7 | 74.2 | 71.3 | 68.8 | 70.2 | 67.5 |
| AkPnBsdf | 78.3 | 80.9 | 75.7 | 92.4 | 92.4 | 92.3 | 69.0 | 69.1 | 69.0 | 60.6 | 60.6 | 60.6 |
| AkPnCGdD | 82.0 | 85.7 | 78.7 | 93.6 | 95.9 | 91.5 | 74.5 | 76.3 | 72.8 | 68.7 | 70.3 | 67.1 |
| AkPnStgb | 75.5 | 83.7 | 68.7 | 92.1 | 90.7 | 93.6 | 69.3 | 68.2 | 70.4 | 62.8 | 61.8 | 63.8 |
| ENSTDkAm | 71.3 | 82.3 | 62.9 | 80.2 | 82.4 | 78.1 | 63.1 | 64.8 | 61.5 | 42.6 | 43.8 | 41.5 |
| ENSTDkCl | 74.4 | 85.8 | 65.8 | 85.6 | 87.5 | 83.8 | 66.1 | 67.5 | 64.7 | 51.4 | 52.5 | 50.3 |
| SptkBGAm | 80.0 | 86.4 | 74.4 | 92.1 | 91.7 | 92.5 | 67.1 | 66.8 | 67.5 | 53.5 | 53.3 | 53.8 |
| SptkBGCl | 81.0 | 86.5 | 76.2 | 92.6 | 93.6 | 91.7 | 74.3 | 75.1 | 73.5 | 59.8 | 60.5 | 59.2 |
| StbgTGd2 | 82.5 | 85.7 | 79.4 | 92.2 | 94.7 | 90.0 | 76.8 | 78.8 | 74.9 | 58.0 | 59.5 | 56.6 |
| **ENSTDkCl30** | **75.8** | **84.7** | **68.5** | **87.3** | **87.4** | **87.2** | 67.6 | 67.6 | 67.5 | 51.4 | 52.5 | 50.3 |
| MAPS D | 72.9 | 84.1 | 64.4 | 82.9 | 85.0 | 81.0 | 64.6 | 66.2 | 63.1 | 47.0 | 48.2 | 45.9 |
| MAPS MIDI | 80.2 | 85.2 | 75.7 | 92.9 | 93.7 | 92.2 | 72.0 | 72.6 | 71.3 | 61.6 | 62.3 | 61.2 |

Table 5. Framewise performance and note-based performance for onsets, offsets and when requiring both onsets and offsets to be correct, on subsets of the MAPS dataset. For *D* and *MIDI*, the performance was computed by averaging the results of each subset. Results for the ENSTDkCl30 subset (bold) are used for comparison with other systems.

Performance is lower for the Disklavier recorded in an ambient setting (ENSTDkAm), and to some extent also for the closely recorded ENSTDkCl. This can partly be explained by the inaccurate annotations of these subsets. They are too early in relation to the sounds; the transmission time from the piano to the microphones is likely unaccounted for. The natural ambience of the recordings may further complicate prediction. A comparison with the systems that have been evaluated for $\mathcal{F}_{on}$ and $\mathcal{F}_{fr}$ on the MAPS ENSTDkCl30 dataset is shown in the top two graphs of Fig. 13. The proposed system just barely outperforms other systems for both tasks. Systems specifically designed for piano transcription (Haw17, SONIC, Kir14, Nak18) generally have the highest results among the comparison systems.

### D. TRIOS

Shown in the middle two graphs of Fig. 13 is the performance comparison for the TRIOS dataset. The upper of these two graphs shows the comparison for note tracking (onsets). The proposed system clearly outperforms other systems, and the error rate is more than halved in comparison with the previous state-of-the-art. Performance for the track "Take Five," containing drums, was rather high ($\mathcal{F}$ = 90.7, $\mathcal{P}$ = 88.7, $\mathcal{R}$ = 92.9). This indicates that the system can be used in regular music mixes with drums. The lower of the two graphs shows the performance comparison for $f_0$ estimation. The system has never analyzed a mix of hammered and sustained instruments during training and struggles to detect offsets, as indicated in Table 4. Many offsets are cut short, which reduces $\mathcal{R}$ for $\mathcal{F}_{fr}$. The system only barely performs a new state-of-the-art.

### E. Woodwind

Shown in the bottom two graphs of Fig. 13 are comparisons for $f_0$ estimation and note tracking performance for the Woodwind track. The music track contains five voices and many tones with staccato. Neither of these features is present in the training set of sustained instruments. Nevertheless, the proposed system outperforms the other systems.

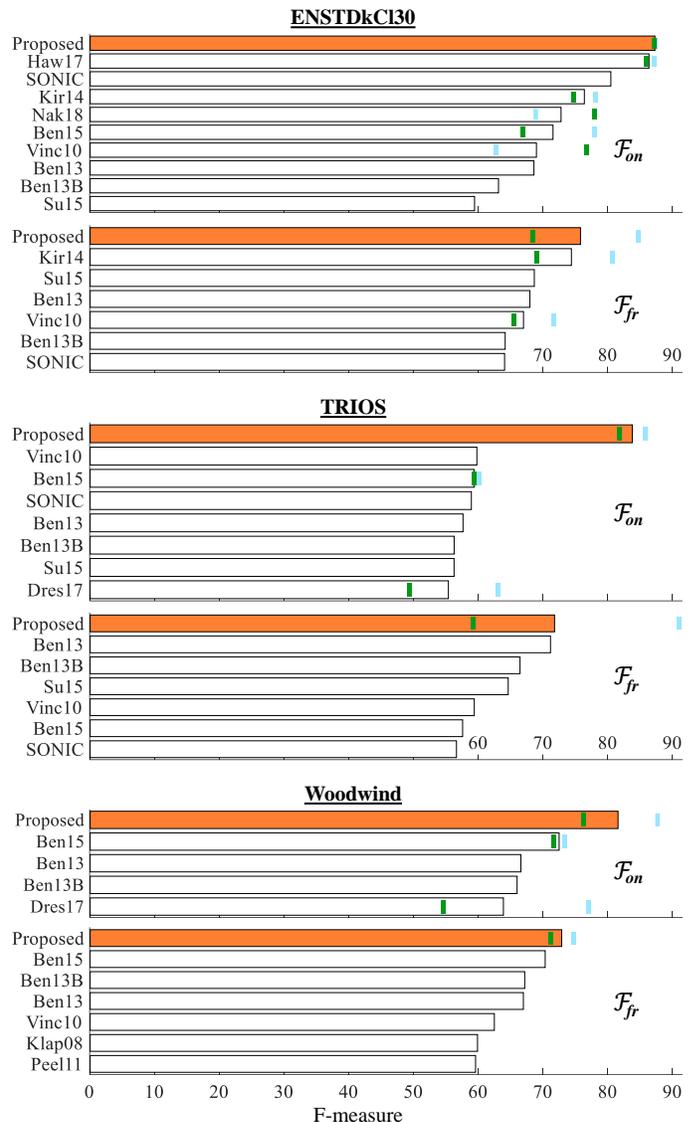

Fig. 13. Comparing $\mathcal{F}_{on}$ and $\mathcal{F}_{fr}$ for: the ENSTDkCl30 subset of MAPS, the TRIOS dataset, and the MIREX Woodwind test track. All results are aligned across the *x*-axis. Dark green horizontal markers represent $\mathcal{R}$ and light blue $\mathcal{P}$.

### F. Combined metric and intermediate evaluation

A single combined metric for performance was computed as the harmonic mean of the four test set results

$$\bar{\mathcal{F}} = \frac{4}{1/\mathcal{F}_{\text{Bach10}} + 1/\mathcal{F}_{\text{MAPS}} + 1/\mathcal{F}_{\text{TRIOS}} + 1/\mathcal{F}_{\text{Woodwind}}}. \quad (10)$$

These final results are reported in Table 6. The table also includes results from Dressler [45]. Furthermore, results at intermediate processing steps are provided to indicate how the system improves from various networks in the DLL architecture [47]. *Evaluation 1* for $\mathcal{F}_{fr}$ are from the thresholding operation described in Section V-D, and for $\mathcal{F}_{on}$ and $\mathcal{F}_{off}$, a similar thresholding was performed optimized on the validation set. *Evaluation 2* is based on the tentative onsets and offsets from Sections VI-VII, with the parameters for smoothing the onset curve (see the Appendix) optimized for performance on the validation set. The methodology from Section VIII-D was applied to the tentative notes to compute a framewise estimate.

| Combined metric $\bar{\mathcal{F}}$ for Bach10, MAPS, TRIOS, and Woodwind | | | | | |
|---|---|---|---|---|---|
| | | *fr* | *on* | *off* | *on+off* |
| Dres17 | | - | 59.1 | 45.0 | - |
| **Proposed** | Evaluation 1 | 75.1 | 38.1 | 32.0 | - |
| | Evaluation 2 | 76.7 | 83.1 | 64.3 | - |
| | **Final Results** | **78.1** | **85.9** | **66.6** | **53.9** |

Table 6. Results using the combined metric $\bar{\mathcal{F}}$ for evaluating the test sets. The evaluation at intermediate steps ("Evaluation 1-2") shows how performance increases throughout the various steps of the DLL architecture.

## XII. CONCLUSIONS AND DISCUSSION

### A. Contributions

This article has suggested novel processing strategies for polyphonic pitch tracking. A method was presented for processing a music spectrogram with the equivalent of a sparse filter kernel, where the relative frequencies of bins that are used and that utilize weight sharing have been learned. Cascading networks using this kernel were applied for $f_0$ estimation, in order to prune the search space efficiently. Tonal contours were extracted to define a framework for onset and offset detection. Since tone fluctuations are negated when extracting input representations relative to the contour pitch, the networks could operate across the contours and process steady and fluctuating tones with the same neural weights. These networks used the "neural flux" as an input feature; the variation of latent representations computed across pitch varying tone contours. The method for computing an offsets position based on cumulative features, classifying each frame as being *before* or *beyond* the offset position, seems to produce accurate output predictions after smoothing and thresholding. By iteratively removing false tentative notes and recomputing affected features, an accurate context could be provided during note classification,

### B. Performance

Given that the system has the highest performance on four different test sets, concerning both $\mathcal{F}_{fr}$ and $\mathcal{F}_{on}$, we conclude that it represents a new state-of-the-art in polyphonic pitch tracking. The fact that the system was trained on a separate training set based on 54 different GM instruments gives reason to believe that it generalizes to a wider variety of music examples than many other systems focused on specific instruments (e.g., using only piano samples as training data). The system also had a high performance for offset detection. Too low precision in tentative notes may degrade performance when spurious note onsets extracted from the *OCS* interact with correct onsets in complex ways. If a correct onset is directly followed by a spurious onset in the same contour, the spurious onset will initially form a note across the contour all the way to the offset. Until the spurious note is removed, the correct note will only consist of a short fragment with an offset at the spurious note onset. This situation highlights the importance of an iterative procedure for removing tentative notes, continually computing an updated context.

### C. Training sets and regularization

The two training sets and the methodology for creating them from MIDI data seem to be useful, given the high performance of the system. A straightforward way of increasing the variation of the training examples, and thereby the overall performance of a system trained by it, would be to use more than one SoundFont or more varied MIDI tracks for training. Another improvement would be to also include real performances to the extent possible. For example, the training set did not contain onsets formed through portamento; such onsets were present in the Bach10 dataset and hard to detect for the system.

The representations extracted during training for one supervised learning step is later used (in some shape) as input for the next learning step. As a result, the subsequent model may become over-reliant on the part of the input that comes from the previous supervised step, and not reliant enough on the data that comes from connections from earlier parts of the system Therefore, if enough data is available or easy to create, a larger pool of training examples can be created than is actually used during each training step. Examples can then be selected randomly for training each learning module. If a large number of instruments or instrument models were used to create the training set, the random sampling could happen on the instrument level. Otherwise, it can happen on the level of the MEs. The random variation of the spectrum used for regularization in this study could also be extended with techniques proposed in [53, 55].

### D. Structural disentanglement and DLL

The system was designed to extract the inherent organization of pitched music at intermediate layers. Tonal contours were extracted so that networks could operate across them for onset and offset detection, and tentative notes were extracted to model further processing around them. This type of "structural disentanglement" promotes invariant processing in two ways [47]: (1) The extracted intermediate representations (contours, notes) are highly invariant with regards to inessential variations in the data, such as spectral variations stemming from instrumentation and performed dynamics [49]. (2) Intermediate representations (e.g., contours and notes) can also facilitate higher-level invariant processing. As shown, the input space can be transformed according to the musical structure. The importance of invariant (or equivariant) representations and their connection to parameter sharing in deep learning is outlined in [56].

Extracted structures can be very beneficial when expanding the system beyond multiple pitch tracking. Pitchogram-like representations have, for example, been used for beat tracking [57] and instrument recognition [58].

# Appendix

## III-B. Filtering the spectrogram

To make sure that the $f_0$ estimation could be run in real-time, for example, during a live performance, the filtering was designed so that it could be applied in a time-causal manner with a relatively small delay (look-ahead) of 0.093 s. However, since the subsequent note-tracking steps are not time-causal, the $f_0$ estimation was not run as such during evaluation.

### 1) Spectral variation ($V^s$)

The spectral variation was computed both across frequency and time in a time-causal manner. Let $M_k$ represent a spectrum vector from $M$ of length 518 at a frame in time. The spectral variation vector of the same shape, denoted by $\tilde{V}_i^s$, was then computed at frame $i$ by taking the mean magnitude of all previous frames

$$\tilde{V}_i^s = 20 \log_{10} \frac{\sum_{k=1}^{i} M_k}{i}. \tag{A1}$$

To reduce the number of operations required, this was computed from the cumulative sum in each frame $M_k^c$ of the magnitude spectrogram

$$M_1^c = M_1, \quad M_k^c = M_{k-1}^c + M_k^c, \quad \tilde{V}_i^s = 20 \log_{10} \frac{M_i^c}{i}. \tag{A2}$$

After this, each frequency bin of $\tilde{V}_i^s$ was smoothed across frequency, using a Hann window with a height of 33 bins, centered at frame $i$. Finally, each frame was normalized with respect to the level, and the spectral variation reduced

$$V_i^s = \frac{\tilde{V}_i^s - \max(\tilde{V}_i^s)}{3}. \tag{A3}$$

### 2) Level variation ($V^l$)

The signal level variation of the music across time was computed from the maximum magnitude $M^{max}$ of each time frame $i$, (across the 518 frequency bins $j$) and the cumulative maximum $M^{cmax}$:

$$M_i^{max} = \max_{1 \le j \le 518} (20 \times \log_{10} M_{ij}) \tag{A4}$$

$$M_1^{cmax} = M_1^{max}, \quad M_i^{cmax} = \max(M_{i-1}^{cmax}, M_i^{max}). \tag{A5}$$

However, for the evaluated model (non-time casual), $M_1^{cmax}$ was replaced with the maximum across the whole track. From a listener's perspective, the cumulative maximum gives a (time-causal) indication of a reasonable playback volume for the track, whereas the framewise maximum indicates the local loudness of the track. The two variables $M_i^{max}$ and $M_i^{cmax}$ were combined into $\tilde{V}_i^l$, a compromise between local and global (up to time frame $i$) considerations, where H represents smoothing across time with a Hann window of size 33

$$\tilde{V}_i^l = H(\max\{M_i^{max}, M_i^{cmax} - 30\}). \tag{A6}$$

The subtraction of 30 means that the vector will rely on $M^{cmax}$ only for fainter parts of the music where $M^{max}$ drops 30 dB below the overall cumulative maximum. From a listener's point of view, $M^{cmax} - 30$ therefore specify a threshold based on the overall listening volume. The output of Eq. A6 was combined with a time-shifted copy of the cumulative maximum

$$V_i^l = \frac{\tilde{V}_i^l + M_{i+16}^{cmax}}{2} - 6, \tag{A7}$$

to form $V^l$. A reason for applying Eq. A7 is to avoid an increase of $V^l$ in tandem with the increased loudness from the first tones (or from the first loud section) of a track. By increasing $V^l$ before the increase in loudness of the track, onsets will not be disguised by a change in the noise floor. Of course, the signal level of the complete track could instead have been established beforehand, but a time-causal implementation was desired for future extension.

### 3) Combining $V^s$ and $V^l$

The noise floor $V^{ls}$ was computed for each time frame $i$ by multiplying the vector $V_i^s$ with the scalar $V_i^l$.

## IV. (C-D) Bin distribution of the learned kernel

The bin distribution and associated learned weights of the sparse kernel is specified in Table A1, together with the constant term. They can be used in combination with the whitening vector $W$ (Section IV-E), available by contacting the author, to compute the Tentogram.

| Bin index | Computed weight | Bin index | Computed weight | Bin index | Computed weight |
|---|---|---|---|---|---|
| -705 | -0.0207 | 9 | 0.0392 | 534 | 0.0119 |
| -655 | 0.0018 | 25 | -0.0268 | 535 | -0.0226 |
| -631 | -0.0109 | 133 | -0.0236 | 557 | 0.0491 |
| -624 | 0.0291 | 217 | -0.0227 | 593 | -0.0559 |
| -601 | 0.0072 | 240 | 0.0415 | 620 | -0.0053 |
| -559 | -0.0511 | 293 | -0.0513 | 674 | 0.0612 |
| -430 | -0.1713 | 315 | -0.0447 | 720 | 0.0051 |
| -429 | 0.1418 | 327 | -0.0094 | 732 | -0.0192 |
| -407 | 0.0079 | 333 | -0.0063 | 738 | -0.0253 |
| -388 | -0.0582 | 380 | 0.0439 | 761 | 0.0046 |
| -324 | -0.0020 | 434 | -0.0769 | 797 | -0.0058 |
| -238 | -0.0961 | 435 | 0.0045 | 802 | 0.0064 |
| -159 | 0.0091 | 448 | -0.0213 | 830 | 0.0742 |
| -142 | -0.0380 | 480 | 0.0192 | 874 | 0.1063 |
| -127 | 0.0043 | 497 | -0.0103 | | |
| -117 | -0.0054 | 505 | 0.0128 | **Constant** | |
| -72 | -0.0313 | 506 | -0.0451 | -5.2314 | |
| 0 | 0.1152 | 520 | -0.0209 | | |

**Table A1.** The active bin indices in the sparse kernel and the associated computed weights. The constant term computed by the network is also included.

## VII. (A-C) On set Detection

Subsequent sections of the Appendix will provide a set of time indexing positions at which features were extracted for each training example. Just as for the frequency bin indices, these can also be understood as filter kernels, if comparing the processing to that of a CNN.

The input to $N_3$ consisted of activations from previous networks and signal levels from the filtered spectrogram. For the output activations of $N_2$, the activations at $\pm 40$ frames were included, using a step size of two (discarding every other frame). This time indexing/filter kernel will be referred to as $T_2$ in the rest of the article. For the hidden layer activations (the "neural flux"), a slightly smaller context window $T_1$ was used, corresponding to time frames (in relation to the evaluated frame) of {-13 -8 -4 -2 0 2 4 8 13}. The fluctuations were computed as the difference between each frame in $T_1$ for each individual neuron. Furthermore, the original activations of the evaluated time frame (frame 0) were also included, providing context about the raw values at which activations changed.

The SF was computed using frequency bins relative to the contour frequency between -186 (down three octaves and a semitone) and +306 (up five octaves and a semitone), using every other bin. This frequency index will be referred to as $F_1$. To also account for information at skipped bins, the spectrum was first smoothed across frequency with a vector of length three and weights {0.25 0.5 0.25}. Relative time indexes {-12 -8 -4 0 4 8} were used ($T_4$), computing the difference over every other index (i.e., the SF between frames -12 and -4, between -8 and 0, etc.). Furthermore, the signal level at the evaluated frame (frame 0) was also provided.

The absolute pitch flux (PF) was computed for the contour $P$, at all time frames $i$ of $T_2$ by

$$|P_{i+1} - P_i|. \tag{A8}$$

The change (derivative) of the $V^l$ was used across indexes $T_2$.

### VII-E. Peak-picking onsets

The first step of the filtering was a thresholding operation, applied on each frame $x$ of the $OC$ by using a smooth thresholding function:

$$x = x - z - r \tag{A9}$$

$$x = \begin{cases} r\left(\frac{e^x}{r} - 1\right), & x < 0 \\ x, & x \geq 0 \end{cases} \tag{A10}$$

$$x = x + r \tag{A11}$$

The threshold $z$ was -4.8, and $r$, stipulating the range within which the threshold smoothly takes effect, was 1. After this, the $OC$ was smoothed with a Gaussian, using $\sigma = 2.8$. Finally, all peaks above 1.2 were extracted as onsets. Recall ($\mathcal{R}$) and precision ($\mathcal{P}$) (see Section X-C) of the onset detection was computed for all combination of the four parameters. The function

$$S = 100\,\mathcal{R} + 3.5 \tan(2\mathcal{P} - 1) \tag{A12}$$

was then applied to the result of each combination, assigning much higher importance to $\mathcal{R}$ than $\mathcal{P}$. This ensures a high recall in the onset detection; the precision was improved at a later stage when a sufficient context for each note had been collected. The tangent function acts as an inverted $s$-function in the applicable range, discouraging too low precisions (below around 0.25). The included factors (100 and 3.5) further balance the importance between $\mathcal{R}$ and $\mathcal{P}$. The parameters that maximized $S$ were used, as previously presented.

### VII. Offset Detection

For computing the offset detection activation, the annotations associated with each correctly detected onset were used to set annotated offset positions. If another annotated offset of the same pitch ($\pm 50$ cents) occurred after a detected onset position but before its corresponding annotated offset position, that note was discarded from the training set. Finally, if a new annotated onset of the same pitch started within four frames of the annotated offset, the last four frames of that note were discarded from the training set. This was done to discourage the network from inferring offset positions by trying to find subsequent onsets.

Features for offset detection were computed as the cumulative activations/magnitudes for the note along time, while also providing the present activation/magnitude. By doing so, the network is provided with information of the characteristics of the note so far, in relation to the present state. The futures for each time frame were:

- The neural activations of the last hidden layer of the pitch network, also using their cumulative mean.
- The output activation of the pitch network at relative time frames given by $T_1$, and a feature consisting of its cumulative mean (up to each time frame).
- The offset curve, $OFC$, at relative time frames given by $T_1$, and a feature consisting of the cumulative sum of these activations. Furthermore, two more cumulative sums were computed by first subtracting 0.1 and 0.2 from the $OFC$, thresholding at 0.
- The pitch kernel signal levels ($K_L$) used as input to the tentogram and pitch network, and a feature consisting of their cumulative mean.
- The pitch and the frame index within each note and region (starting from 1).

To determine the final offset position, the output activations were smoothed across time with a Gaussian window ($\sigma = 4.3$), and the first time-frame with an activation $a > 0.47$ was chosen as the offset. These two parameters ($\sigma$ and $a$) minimized the absolute distance to the annotated offset in a grid search across the training set.

### VIII. Note Classification

The following information from the onset of the note was used for the input layer:

- The previously computed $OC$ indexed at $T_2$.
- The last hidden layer of $N_3$, indexed at $T_{4*}$.
- The input to the pitch network, indexed at $T_{4*}$.
- The signal level of $L_{+25}$, indexed at $F_1$ and $T_4$.
- The flux of $V^L$ (Section III-B), using relative frames between -15 and 15 with step size 3.

In order to provide variation in input, the time indexing of $T_4$ was shifted two frames forward to be {-10 -6 -2 2 6 10} ($T_{4*}$).

Order statistics in the form of quintiles {25 50 75}, were collected for activations and signal levels across the length of the note for: the PF, the absolute PF, the *OC*, the *OCS*, the activations of the pitch network at the output layer and last hidden layer, and the 50 input features from $L^4$ to the tentogram network collected at the frequency index closest to the contour pitch.

Other features were computed based on the relationships between each note and adjacent notes extracted from the *same region*. These features included:
- The duration, inter-onset interval (IOI) and inter-pitch interval (IPI) of the evaluated note, the two previous notes, and two subsequent notes. If surrounding notes did not exist in the region, the duration was set to two seconds, the IOI to 3 seconds, and IPI to 0. The last note of the region was assigned an IOI one second longer than its duration, and an IPI of 0.
- The order of the evaluated note in the region, as well as the number of notes extracted from the region, were included as two additional features.

Furthermore, the onset activation of close onsets was collected, by summing the activations at each semitone relative to the evaluated note at $\pm 25$ semitones. The activations were summed within five time-ranges relative to the onset of the processed note {-18 to -11; -10 to -4; -3 to 3; 4 to 10; 11 to 18}.

The threshold for when to stop the note removal (Section VIII-B) was set to 0.55, determined in a small grid search on the training set.

During the synthetization and annotation of the *training* set, the onset times of *training* annotations were shifted forward based on the shape and length of the attack (envelope) of each instrument (see Section X of the Appendix below). Test sets are however oftentimes (arguably incorrectly) annotated based on the time of the first sound produced in a tone. To account for this, the predicted note onset times were shifted 0.01 s earlier after the note classification step *during run-time*. Annotations in the *test* sets were not adjusted.

## VIII. Regularization

All time frames of $M$ was multiplied by a vector $\mathcal{G}$, spanning the 518 frequency bins. This vector was derived from the equivalent decibel scale vector $\mathcal{G}_{dB}$, which was created by summing smooth filters. First, ten Hann windows were used. The width $w$ of each window was drawn from the uniform random discrete distribution $U_d$ between odd integers 1-241

$$w \sim U_d([1, 3, 5, \ldots, 241]) + 120, \quad (A13)$$

and the center $c$ of the window across $F$ was randomly selected from 518 potential integer values

$$c \sim U_d([1 - 518]). \quad (A14)$$

The uniform random distribution $U$ was used to select the amplitude $a$ of each window as a floating-point number between -7 and 7,

$$a \sim U([-7 - 7]). \quad (A15)$$

The Hann windows designed with the parameters in Eqs. A13-A15 were summed to $\mathcal{G}_{dB}$, ignoring any parts of a window that fell outside the boundaries of the vector. A shelf window was then added with amplitude drawn from

$$U([-3.5, 3.5]), \quad (A16)$$

and a center $c$ determined by applying Eq. A14. The shelf fell or rose to 0 linearly over 60 bins, centered at $c$.

Finally, the filter $\mathcal{G}$ was derived by subtracting the mean, and converting from the log-domain.

## X. Datasets and Evaluation Method

When a MIDI note starts, the corresponding audio samples have an attack during which the energy rises. The perceived onset time of the note will depend on the shape and length of that attack. When a MIDI note ends, the corresponding audio samples have a decay during which the energy falls. The perceived offset time of the note will depend on the shape and length of that decay. Therefore, to increase the accuracy of the onset and offset annotations in the *training* set, the approximate delay between the MIDI onset/offset and the perceptually correct onset/offset was estimated by the author for each general MIDI (GM) instrument (GMI) of the SoundFont (manually). The annotated onset times were as a result shifted forward in time 0-0.09 seconds (s) (mean 0.008 s) for plucked and hammered instruments and 00.093 s (mean 0.024 s) for sustained instruments, relative to the MIDI onset information. The annotated offset times were shifted forward in time 0-0.6 s (mean 0.10 s) for plucked and hammered instruments and 0-1 s (mean 0.18 s) for sustained instruments. Annotations for each note in the training set were updated by accounting for the delay previously determined for each instrument.

### XI. Evaluation

Training set performance can give an insight into, e.g., the amount of overfitting during training in relation to test set performance, and is therefore provided here. The $\mathcal{F}_{fr}$ (Section X-C) for the *Attacked* and *Sustained* datasets was 88.9 and 95.6 respectively, whereas $\mathcal{F}_{on}$ was 96.4 and 97.1 respectively. The recall $\mathcal{R}$ of potential onsets was kept high initially to retain as many as possible for later classification, while pruning away false examples and increasing $\mathcal{P}$. It was {99.8, 99.6, 97.9} at the output of networks $N_{1-3}$, on average across the two datasets. For this statistic, $\mathcal{R}$ was measured in the Tentogram and Pitchogram as the proportion of onset annotations within 50 cents and 120 ms of an activated $f_0$.